\documentclass[namedreferences]{solarphysics}
\usepackage[optionalrh,solaromanenum]{spr-sola-addons} 
\usepackage{graphicx}                    
\usepackage{color}                       
\usepackage{url}                         

\usepackage[pdfborder={0 0 0 },urlcolor=blue,breaklinks]{hyperref}
\ifx \doiurl    \undefined \def \doiurl#1{\href{http://dx.doi.org/#1}{\textsf{DOI}}}\fi
\ifx \adsurl    \undefined \def \adsurl#1{\href{http://adsabs.harvard.edu/abs/#1}{\textsf{ADS}}}\fi
\ifx \arxivurl  \undefined \def \arxivurl#1{\href{http://arxiv.org/abs/#1}{\textsf{arXiv}}}\fi

\newcommand{\ms}{\rm{\,m\,s^{-1}}}

\newcommand{\Mm}{\rm{\,Mm}}

\newcommand{\secs}{\rm{\,seconds}}
\newcommand{\mHz}{\rm{\,mHz}}

\begin{document}

\begin{article}

\begin{opening}

\title{Additional Evidence Supporting a Model of Shallow, 
High-Speed Supergranulation}


%
\author{T.L.~\surname{Duvall Jr.}$^{1}$\sep
        S.M.~\surname{Hanasoge}$^{2,3}$\sep 
	  S.~\surname{Chakraborty}$^{4}$
      }


%
\runningauthor{Duvall \emph{et al.}}
\runningtitle{Shallow High-Speed Supergranulation}

%
\institute{$^{1}$ Solar Physics Laboratory, NASA Goddard Space Flight Center,
Greenbelt, MD, 20771, USA
email:\href{mailto:Thomas.L.Duvall@nasa.gov}{\sf{Thomas.L.Duvall@nasa.gov}}\\
$^{2}$ Tata Institute of Fundamental Research, Mumbai 400005, India\\
email:~\url{hanasoge@tifr.res.in}\\
$^{3}$ Max-Planck-Institut fur Sonnensystemforschung, Justus-von-Leibig-Weg 3, 37077 G\"{o}ttingen, Germany\\
$^{4}$W.W. Hansen Experimental Physics Laboratory, Stanford University, Stanford, CA 94305, USA
email:~\url{deepc@stanford.edu}
}



\begin{abstract}
Recently, Duvall and Hanasoge ({\it Solar Phys.} {\bf 287}, 71-83, 2013)
found that large distance $[\Delta]$ separation
travel-time differences from a center to an annulus
$[\delta t_{\rm{oi}}]$ implied a model of the average supergranular cell
that has a peak upflow of $240\ms$ at a depth of $2.3\Mm$ and a 
corresponding peak outward horizontal flow of $700\ms$ at a depth of
$1.6\Mm$.  In the present work, this effect is further studied by 
measuring and modeling
center-to-quadrant travel-time differences $[\delta t_{\rm{qu}}]$,
which roughly agree with this model.  Simulations are analyzed that
show that such a model flow would lead to the expected travel-time 
differences.  As a check for possible systematic errors, the 
center-to-annulus travel-time differences 
$[\delta t_{\rm{oi}}]$ are found not to vary with heliocentric angle.
A consistency check finds an increase of 
$\delta t_{\rm{oi}}$ 
with the temporal frequency $[\nu]$ by a factor of two,
which is not predicted by the ray theory.
\end{abstract}

%

\end{opening}

%

\section{Introduction}

Supergranulation, first seen as a 30 Mm cellular pattern of horizontal 
flows detected by Doppler shifts \cite{Hart54,Leighton62} in the 
solar photosphere, continues to puzzle investigators 
(see review by \opencite{Rieutord10}).  
Recent work attempts to understand supergranulation by revealing
its subsurface structure by numerical simulations \cite{Stein06a}
or by local helioseismology \cite{Gizon10}.

Detailed radiative-hydrodynamic simulations of the outer convection 
zone and atmosphere show no excess flow signal 
at the supergranular scale in the photosphere, in contrast
to the observational results \cite{Nordlund09}.
These simulations, which match the observations of the solar granulation
so well, would seem to have all of the ingredients required to reproduce 
supergranulation.  In particular, the early suggestion of 
\inlinecite{Leighton62} that
He {\sc II} ionization could give rise to supergranulation, is tested by the
simulations with a null result.  
One possibility remaining to be tested is the simulation of magnetic
field, which is known to be present along cell boundaries.  

Local helioseismology has been used extensively to study supergranulation
(see review by \opencite{Gizon10}), although no consensus has emerged about 
fundamental questions such as the depth of the peak flow and the existence 
or not of counterflows at depth. 
Some efforts centered on making inversions of individual realizations
of the supergranular flow field 
{ \bf
\cite{Duvall97,Zhao03,Woodard07,Jack08,Svanda11}.  
}
In some of the work there is great difficulty
in separating a horizontally diverging outflow from an upflow
\cite{Zhao03,Dombroski13}, 
although in other work this may have been solved 
{ \bf
\cite{Svanda11}.
}
To make flow maps of individual supergranular realizations,
it has been necessary to restrict the measurements to small separations
$[\Delta < 5^\circ]$ for which the signal-to-noise ratio is large.  

To measure the general properties of supergranulation, a large number of
cells needs to be examined (in the present work, $6\times10^4$ supergranules
are analyzed).  To increase the signal-to-noise ratio (S/N), 
spatial averages are made about cell locations determined from shallow
signals such as peaks in the flow divergence.  Such a method was first
used by \inlinecite{Birch06} and subsequently by \inlinecite{Duvall10}
and \inlinecite{Svanda12}.
Weak signals can be separated cleanly from realization noise, although 
more attention to systematic errors is required.
As noticed by \inlinecite{Svanda12}, the present method of defining
cells is probably biased towards larger cells than the average.  
This might be corrected (in the future) by directly modeling the
spatial autocovariance of the travel-time maps.

\begin{figure}
\centerline{\includegraphics[width=1.0\textwidth,clip=]{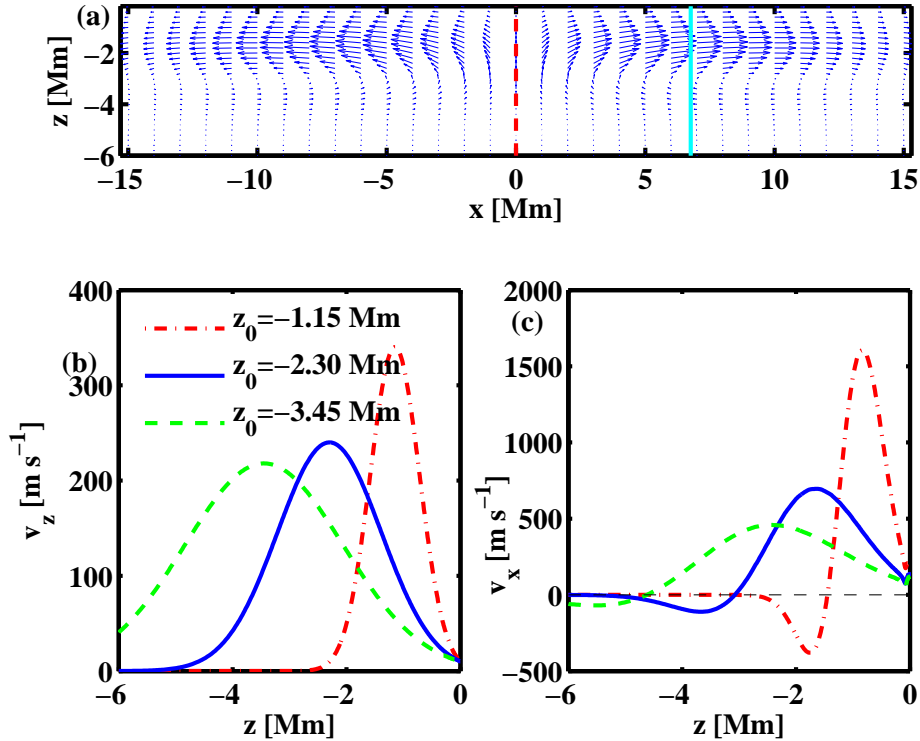}}
\caption{
The flow models from Article I.  (a) Velocity vectors for the best model.  This
is the model labeled {\sf g2} in Table 1 of Article I, with peak upward flow of 
$240\ms$ at $z=-2.3\Mm$ and peak horizontal flow of $700\ms$ at 
$z=-1.6\Mm$ and $x=7\Mm$.  The cuts shown in (b) are taken at the location
of the red dashed vertical line in (a).  
The cuts in (c) are taken at the location of the
turquoise line in (a) at $x=7\Mm$.  
(b) Cuts of the vertical flow at cell center for
the three models in Article I, model {\sf g1} (green; dashed), 
model {\sf g2} (blue; solid), 
and {\sf g3} (red; dot-dashed).
(c) Cuts of the horizontal flow versus height at the location of the peak
flow.  Colors and line styles are the same as in (b).  
}
\label{F-flowmodels}
\end{figure}

The averaging of the signals from many cells makes it possible to use
larger $\Delta$s (up to $24^\circ$ in the present study), which would 
normally not be feasible for a 12-hour observation because of the 
increased noise due to the amplitude reduction from the geometrical
spreading of the wavefront \cite{Gizon04}.
The separation of
the horizontal and vertical flow signals is much better at larger $\Delta$,
as the rays are more vertical in the critical near-surface region.
\inlinecite{Duvall13} (hereafter Article I) found that 
the center-to-annulus travel-time difference $[\delta t_{\rm{oi}}]$ was 
roughly constant at $5.1$ seconds in the range $\Delta=10-25^\circ$.
In a simple ray-theory interpretation, 
this requires a vertical upflow considerably larger than the $10\ms$ 
observed at the photosphere \cite{Duvall10} and in fact the best-fit model
had a peak upflow of $240\ms$ at $z=-2.3\Mm$. 
Plots of this model and the bracketing models are shown in 
Figure~\ref{F-flowmodels}.
That large vertical upflows are required was recently confirmed by
the analysis of \inlinecite{Svanda12} by a considerably different formalism.

The strategy for obtaining the best model was developed in Article I and 
is as follows:
We assumed the
simplest vertical-flow model that reduces to a $10\ms$ vertical flow at the
surface and still approaches the $5.1$ seconds for the asymptotic behavior 
of the $\delta t_{\rm{oi}}$ signal at large $\Delta$.  
This is the gaussian with a single peak.
For a particular choice of depth of the peak vertical flow $[z_0]$, 
the width of the
gaussian and its amplitude are determined uniquely  by the $5.1$ seconds
$\delta t_{\rm{oi}}$ signal
requirement and the $10\ms$ upward flow at the photosphere.  With some
reasonable choices for the horizontal parameters $k$ and $R$ (see Article I),
the horizontal flow is then determined from the vertical flow and the
continuity equation.  Three models were examined that bracket the
observations.  These are distinguished by the height of the peak flow,
$z_0=-1.15\Mm$, $z_0=-2.30\Mm$, and $z_0=-3.45\Mm$.  
The $\delta t_{\rm{oi}}$ signal 
is computed from
the ray theory using both the vertical and horizontal flow components.
We found that the $z_0=-2.30\Mm$ model was most similar to the observations.
For the $z_0=-3.45\Mm$ model (and any with a deeper $z_0$), the horizontal
component contributes significantly and leads to a behavior at large
separations that is inconsistent with the observations.  We conclude that
if there is a deeper horizontal flow, it must have a small magnitude to
not be observed in the $\delta t_{\rm{oi}}$ signal.

In the present work, the efforts of Article I are extended to include 
quadrant analysis in Section~\ref{sec-quad}, an attempt to measure a
heliocentric-angle (or center-to-limb) dependence in Section~\ref{sec-heliocen},
tests with simulations in Section~\ref{sec-sim},
and an attempt to measure a temporal-frequency $[\nu]$ dependence in
Section~\ref{sec-nu}. We give some conclusions in Section~\ref{sec-dis}.

\section{Analysis}

\subsection{Quadrant Analysis}
\label{sec-quad}
\begin{figure}
\centerline{\includegraphics[width=0.9\textwidth,clip=]{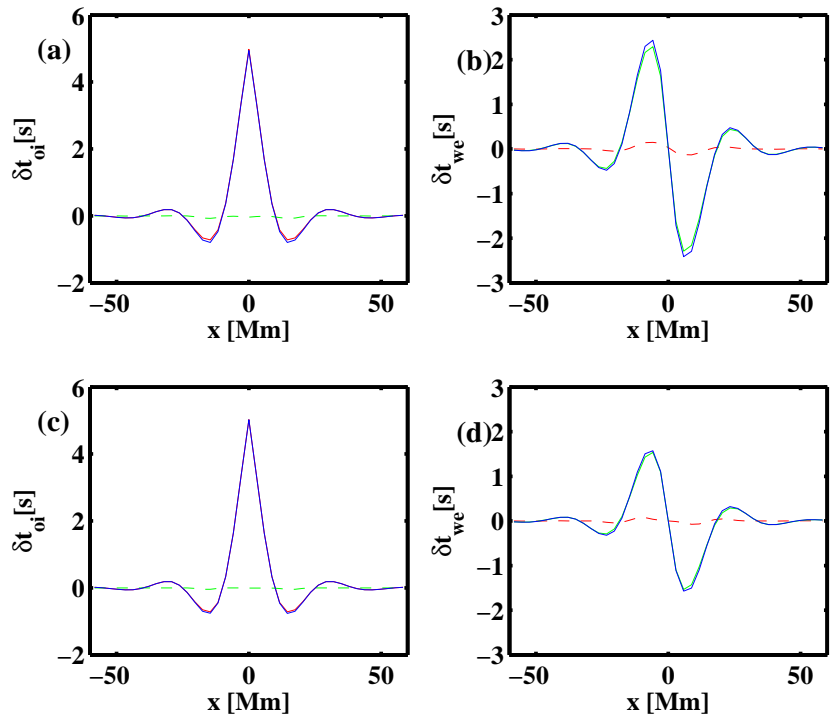}}
\caption{
Cuts in the east--west direction across model maps for 
center--annulus travel time differences $[\delta t_{\rm{oi}}]$ ((a) and (c))
and west--east travel-time differences $[\delta t_{\rm{we}}]$ ((b) and (d)) for
distances $\Delta=11.76^\circ$ ((a) and (b)) and $\Delta=20.64^\circ$ 
((c) and (d)).
Vertical-flow contributions to the travel times are in red
(solid in (a) and (c); dashed in (b) and (d); horizontal 
contributions are in green (dashed in (a) and (c); solid in (b) and (d)); 
and the sum is in blue.  The model is the 
nominal one from Article I ({\sf g2}) in which the peak upward vertical 
flow is at 
a depth of $2.3\Mm$ with magnitude of $240\ms$.  The peak horizontal flow
is at a depth of $1.6\Mm$ at a distance $7\Mm$ from cell center and with
velocity $700\ms$.
In general it is very difficult to separate the sum signal (blue) from
the dominant signal (vertical flow (red) in (a) and (c) and 
horizontal flow (green) in (b) and (d)).
}
\label{F-model_cuts}
\end{figure}

In Article I
it was shown that the center--annulus travel-time 
difference $[\delta t_{\rm{oi}}]$ at large distances $[\Delta]$ 
is mostly sensitive to the supergranular vertical-flow signal.  
We might expect that at large $\Delta$ that the center-to-quadrant
signals $[\delta t_{\rm{qu}}]$, where $qu$ corresponds to 
the West--East and North--South quadrant signals $[\delta t_{\rm{we}}$
and $\delta t_{\rm{ns}}]$, would be mostly sensitive to the horizontal
supergranular flow.
To test this idea, travel-time difference maps were constructed with
ray-theory modeling of the average supergranule-flow model {\sf g2} 
from Article I.
The results are shown in Figure~\ref{F-model_cuts} with the horizontal,
vertical, and sum flow contributions to the travel time differences shown 
separately.  For these relatively large $\Delta$s of $11.76^\circ$ and
$20.64^\circ$, the center--annulus time differences $[\delta t_{\rm{oi}}]$
show very little contribution to the peak signal
from the horizontal flow (relative magnitude 0.008 for 
$\Delta=11.76^\circ$ and 0.001 for $\Delta=20.64^\circ$).  
The contribution of the vertical signal to the peak $\delta t_{\rm{we}}$
is a little larger (0.064 for $\Delta=11.76^\circ$ and 0.050 for
$\Delta=20.64^\circ$), but still small.  Therefore the center--annulus
differences $[\delta t_{\rm{oi}}]$ and quadrant directional differences
$[\delta t_{\rm{qu}}]$ do separate the vertical and horizontal contributions
quite well, if the ray theory can be believed.
It would be very difficult to construct a model with these 
responses in which the 
horizontal flows are leaking into the $\delta t_{\rm{oi}}$ signal to 
yield the five-second signal at large $\Delta$.

To compare the quadrant signals with the models, Helioseismic and Magnetic
Imager (HMI) data were analyzed as in Article I with some improvements.  
The same 64 12-hour intervals 
(10 June 2010-10 July 10 2010) were
analyzed with the same constant
degree--width filter with width $\Gamma_{\rm{\ell}}=400$.  
Cross-correlation
maps were constructed for each 12-hour period for the in and out-annulus
signals and for the four quadrant signals (eastward, {\it etc.} 
direction of waves)
for the 14 distance ranges of Article I and two additional ones at small 
$\Delta$ (centered at $1.20^\circ$ and $1.44^\circ$).
The coordinate system used has equal increments in longitude and
latitude of $0.24^\circ$.  
Gizon--Birch travel times \cite{Gizon04} were
computed for each set of cross correlations and the desired differences:
$\delta t_{\rm{oi}}$, $\delta t_{\rm{we}}$, and $\delta t_{\rm{ns}}$.
The reference correlation was taken as the average over the in and out
correlations averaged over the map, which is of size $96.24^\circ$
on a side.  
An average of these travel-time differences is made about the 
supergranular centers.  In this average, the latitude--longitude travel-time
difference 
maps are transformed locally to a Postel's coordinate system centered on the 
feature.
In this way, features at different latitudes are treated equally and the
resultant average maps can be compared more readily with theory.
Note that this was not the case in Article I, where the averages about the 
feature locations were done in the latitude--longitude coordinate system.
However, as only the center of the maps where the peak $\delta t_{\rm{oi}}$
signal was obtained were used, it was acceptable.
The supergranular-averaged 12-hour maps are averaged over the 64 different
intervals.  One advantage of doing the analysis in this way is that an estimate
of the error can be made at each map location from the scatter of the 64
different intervals.  One might imagine that the scatter would be larger
where the average signal is larger just because of the variability in 
the supergranular signal.  However, this was not the case, and the maps
of errors showed no distinguishable features.

Some background signals are removed from these superposed images before further
analysis.  The $\delta t_{\rm{oi}}$ signal approaches a nonzero constant
far from the central feature.  This constant is measured and removed for
each $\Delta$ range from
the overall image as in Article I.  The $\delta t_{\rm{we}}$ signal has a
relatively large constant (six seconds for the smallest $\Delta$ to 
one second for the largest) removed.  This signal is due to the average 
rotation over the field.  
This signal could be reduced by adjusting the tracking rate from the 
nominal Carrington rate.
The $\delta t_{\rm{ns}}$ signal has both constant offsets and
slopes in the North--South direction at different distances $[\Delta]$.  
The magnitudes are generally small ($<$ one second), but they are present in the
background and hence were removed.  Of these signals 
subtracted, only the rotation one is well understood.  

\begin{figure}
\centerline{\includegraphics[width=1.0\textwidth,clip=]{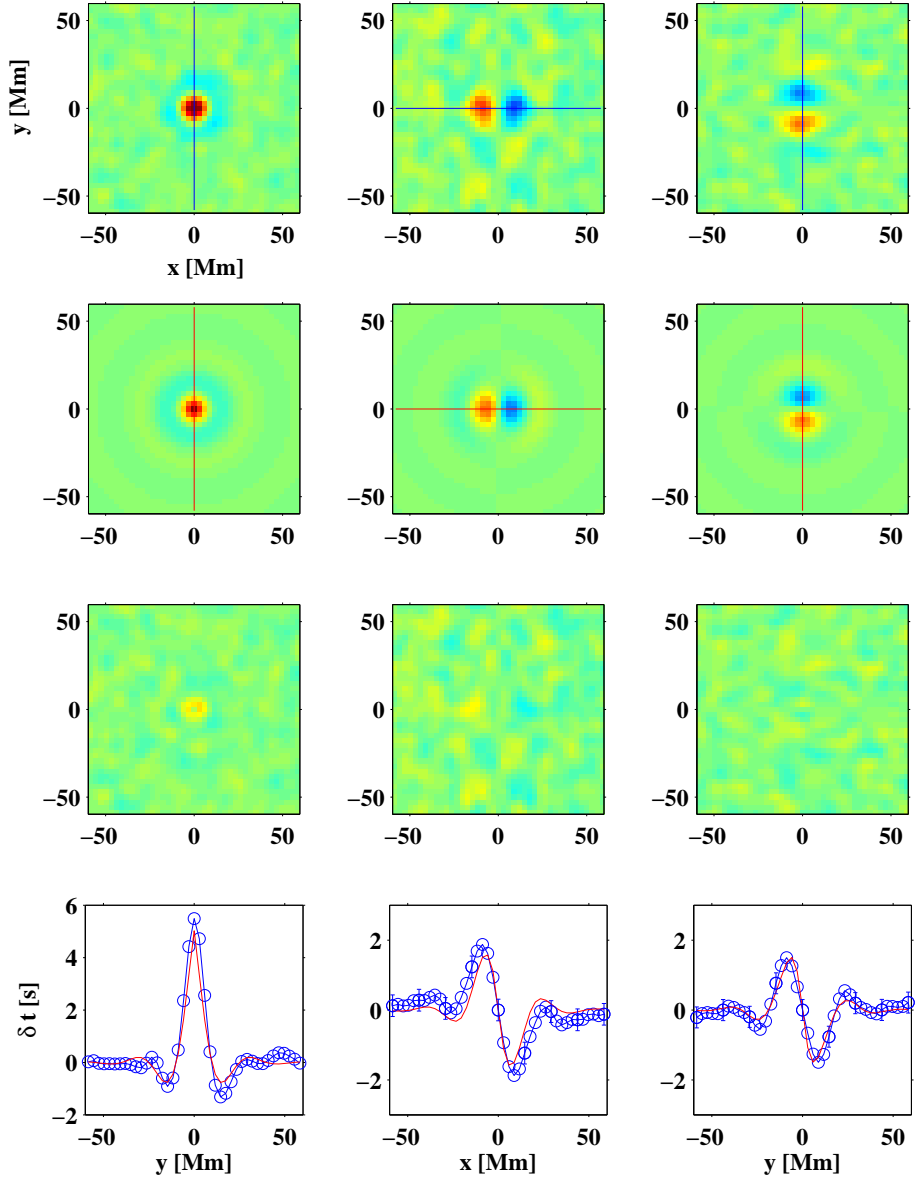}}
\caption{
Comparison of data with models for the distance range 
$\Delta=19.1$\,--\,$22.2 ^\circ$.
The first column is the center--annulus travel time difference
$[\delta t_{\rm{oi}}]$,
The second column is the quadrant West--East travel time difference
$[\delta t_{\rm{we}}]$ and the third column is North--South 
$[\delta t_{\rm{ns}}]$.
The top row is for the data 32-day average of all supergranules.
The second row is the model and the third row is the difference
of data minus model.  The fourth row are cuts through the data and
model images with data shown as blue (with symbols)  and model as red.  
Where the cuts
are taken is shown by the lines across the maps in the
top two rows.  The color scales of the images 
are in a spectral sequence with the first column covering -5 [seconds] 
for blue and +5 [seconds] for red.  This is reduced in the second and third
columns to -3 [seconds] for blue and +3 [seconds] for red.
The dominant blue-green color corresponds to 0 [seconds].
The observed images are made antisymmetric in x about the center point
in the middle column and antisymmetric in y about the center point in
the right column.  
}
\label{F-cmp_model}
\end{figure}

Comparisons between models and data are shown in Figure~\ref{F-cmp_model} for
the 15th (of 16) distance range ($\Delta=19.1$\,--\,$22.2 ^\circ$).  
$\delta t_{\rm{oi}}$ are shown in the left column, $\delta t_{\rm{we}}$
in the middle column, and $\delta t_{\rm{ns}}$ in the right column.
The 32-day average data are shown in the top row; model {\sf g2} from Article I
images are shown in the second row; and the residual (data -- model) are shown
in the third line from the top.  The fourth line shows cuts across 
important parts of the data and model images.  
In general, there is good agreement between the data and the ray
theory modeling.  
The only apparent systematic difference between the data and the model
is in the $\delta t_{\rm{oi}}$ signal near the center.  It would seem
that the width or amplitude of the model could be slightly adjusted.

One might imagine that if the five-second peak signal in $\delta t_{\rm{oi}}$
were due to an incorrect kernel, that the quadrant signals might be very
different from the predicted.  However, this is not the case.  It does not
preclude the possibility that at these large separations all of the
ray kernels are multiplied by some factor.  \inlinecite{Birch07}
have found a case where
the ray kernel is a factor of two larger than the Born-approximation kernel,
although this was at small $\Delta$ and for the fundamental $f-$mode.

\begin{figure}
\centerline{\includegraphics[width=1.0\textwidth,clip=]{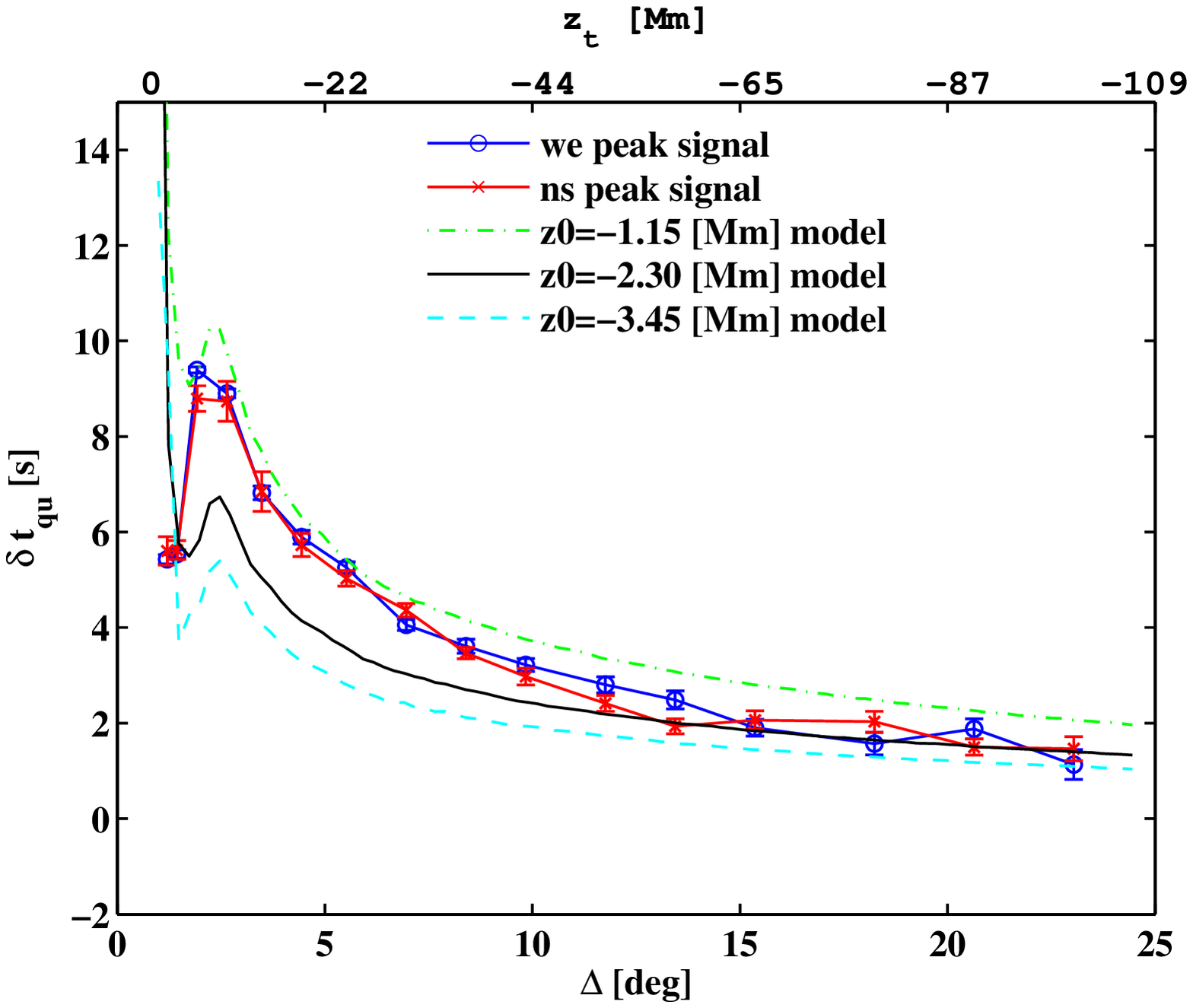}}
\caption{
Measured West--East and North--South quadrant peak travel-time differences 
$[\delta t_{\rm{qu}}]$ versus those from the three models from Article I.  
The blue symbols and
error bars (hardly visible) are for the $\delta t_{\rm{we}}$ and the
red for the $\delta t_{\rm{ns}}$.
The central model (black) is the one that agreed best with the 
$\delta t_{\rm{oi}}$ signal.
$z_0$ is the height of the peak vertical flow.
On the bottom axis is shown the separation between center and 
quadrant $[\Delta]$ and on the top axis the corresponding turning point
depth $[z_t]$.
}
\label{F-wens_vs_model}
\end{figure}

To compare the $\delta t_{\rm{we}}$ and $\delta t_{\rm{ns}}$ signals
with model predictions, the quadrant signals are characterized by
the peak signal in the cuts shown in Figure~\ref{F-cmp_model}.  
Once the maximum signal is found, the neighboring two points are used to
compute a parabola to refine the value of the maximum.  
Model travel-time images are treated the same way as the data.
The three models from Article I are compared with the peak quadrant signals 
in Figure~\ref{F-wens_vs_model}.  At the upper half of the $\Delta$ range,
the observed signals agree pretty well with the model with peak vertical
flow at depth $2.30\Mm$, the same model that agreed best with the 
$\delta t_{\rm{oi}}$ signal in Article I.  At shorter $\Delta$, the
observed signals deviate significantly from that model and agree better
with the more shallow model with peak vertical flow at depth $1.15\Mm$.
Either the precise form of the model is not right or the kernels are 
incorrect.  

\subsection{Heliocentric Angle Analysis}
\label{sec-heliocen}
Recently it has been found that there are flow artifacts with a 
center-to-limb, or heliocentric angle $[\theta]$ dependence \cite{Zhao12}.
As the center--annulus travel time difference $[\delta t_{\rm{oi}}]$
of $5.1 \pm 0.1\secs$ at distances $\Delta > 10^\circ$ is a somewhat 
surprising signal, it was decided
to check whether there is any $\theta$-dependence, which would indicate
an artifact.  This is done by separating
the different supergranules into 11 bins based on the heliocentric angle
of the cell center at the central observation time.  The bins were chosen to
give roughly equal numbers of features in each bin to attempt to equalize
the errors from the different bins.  

The analysis follows the quadrant analysis in Section~\ref{sec-quad},
except that when the average cross correlations are computed about the
supergranular centers, there are 11 averages computed for the different
$\theta$ bins.


\begin{figure}
\centerline{\includegraphics[width=0.9\textwidth,clip=]{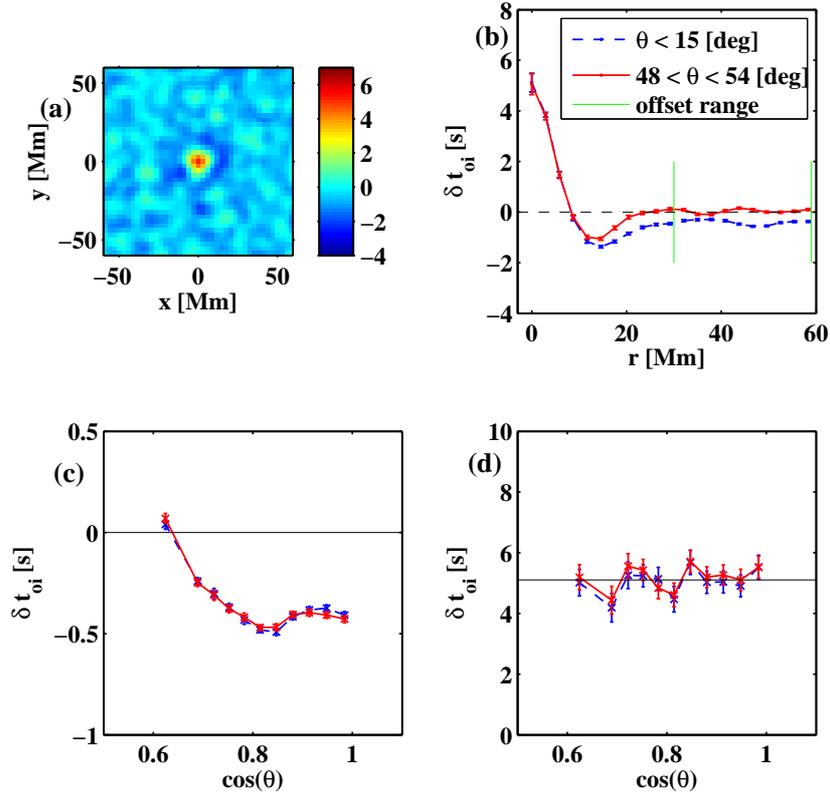}}
\caption{
(a) Center--annulus travel-time difference averaged about supergranule
centers for the range of $\Delta=14.4$\,--\,$24^\circ$ for heliocentric
angle $\theta<15^\circ$.  The scale of the
colorbar at right is in seconds.  
(b) Azimuthal average of (a) (blue; dashed) and for 
$\theta=48$\,--\,$54^\circ$ (red; solid).  
The range $r=30$\,--\,$60\Mm$ over which the
offset is averaged is noted by the vertical green lines.
Note the offset at large radii that is smaller (in absolute value) 
at large $\theta$.
This offset is believed to be an artifact which needs to be removed from
the results.
(c) The offset at $r=30$\,--\,$60\Mm$ for the different travel-time definitions
{\it versus} $\cos(\theta)$.
Blue (dashed) is for the Gabor wavelet phase time differences.  
Red (solid) is for the Gizon--Birch phase time differences.
(d) The resultant travel time differences averaged for the
64 12--hour datacubes corrected for the offset in (c) {\it versus}
$\cos(\theta)$.  The colors are the same as in (c).
In (a)\,--\,(d) distances were averaged over the range 
$\Delta=14.4$\,--\,$24^\circ$.
}
\label{F-tt_rho}
\end{figure}

Gizon--Birch travel times are
computed for each 12-hour interval and the results from the 64 intervals
are averaged.  
The reference correlation was taken as the average over the in and out
correlations averaged over the superposed map, which is of size $9.84^\circ$
on a side.  
There is thus a separate reference correlation for each $\theta$ bin.
This is important as the reference correlation varies from center to limb,
and if it is not taken into account, spurious results are obtained.
With the 11 bins in heliocentric angle, there was insufficient
S/N ratio to compute Gabor-wavelet times for the individual 12-hour intervals.
So, the correlations from the 64 12-hour intervals were averaged and the
Gabor wavelets were fit to the result.  
With the travel-time differences computed in this way, 
the Gabor wavelet and Gizon--Birch
times are almost identical.  This should not be too surprising as the same
correlation windows are fit in the two cases.  Even the noise in the 
resultant travel-time difference maps is highly correlated.

Results for the center--annulus travel-time difference $[\delta t_{\rm{oi}}]$
are presented in Figure~\ref{F-tt_rho}.  The average Gabor-wavelet-phase 
time difference for the first bin ($\theta < 15^\circ$) 
and for the distance range $\Delta=14.4$\,--\,$24^\circ$ is shown in 
Figure~\ref{F-tt_rho}a.  
The normal large positive signal of $\approx$ five seconds is seen at 
cell center
surrounded by a negative moat with signal $\approx -$one $\secs$ corresponding
to the region of downflow of the average cell and also the downflow for
neighboring cells.  The overall mottling of the picture has roughly
supergranular scale and is presumably due to incomplete averaging of 
the supergranular field.  To get a better indication of the average signal,
which we expect to be azimuthally symmetric about cell center, an azimuthal
average of the signal in Figure~\ref{F-tt_rho}a is shown in 
Figure~\ref{F-tt_rho}b (blue;dashed).  Also shown in Figure~\ref{F-tt_rho}b in 
red (solid) is the azimuthal average signal in the outer $\theta$ bin.  

We would expect a zero signal far from cell center.  That it is not, at least
for the inner $\theta$ bin, is apparent in Figure~\ref{F-tt_rho}b.  
This signal, which we measure
as the average on $r=30$\,--\,$60\Mm$, is shown in Figure~\ref{F-tt_rho}c 
for Gabor-wavelet fitting
(blue;dashed) and Gizon--Birch times (red;solid).  
The variation with $\cos{\theta}$ suggests
that this signal is an artifact that could safely be used to correct the
signal at cell center.  It may be related to the center-to-limb flow 
artifact reported by \inlinecite{Zhao12}, as a cell center at disk center
will see a $\delta t_{\rm{oi}}$ of roughly the magnitude shown for the
annulus radii used here.

The results in Figure~\ref{F-tt_rho}c are subtracted from the cell 
center signal to yield the
corrected cell center signal in Figure~\ref{F-tt_rho}d.  
Again the Gabor-wavelet times and the
Gizon--Birch times are very close and in addition, no significant center-to-limb
signal is apparent.  Fitting a line to the results yields a slope with
an error about equal to the value. 
The average over $\theta$, $5.1\secs$, is consistent
with the results of Article I.

\subsection{Simulations}
\label{sec-sim}
\begin{figure}
\centerline{\includegraphics[width=1.0\textwidth,clip=]{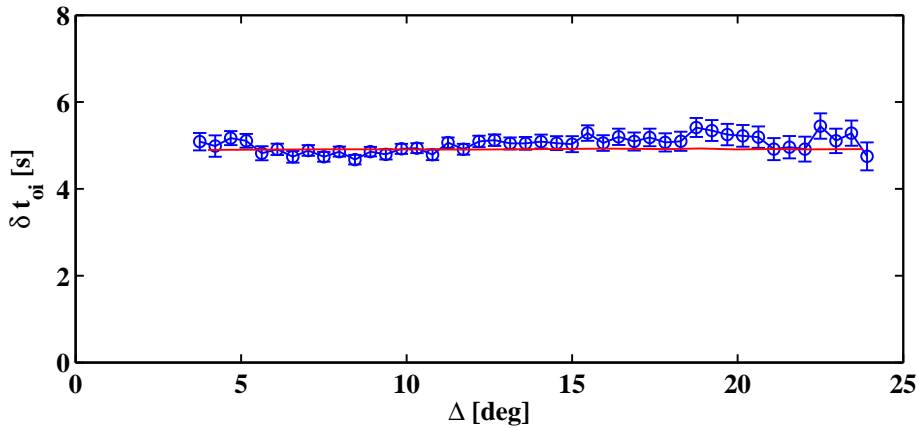}}
\caption{
Comparison of the center--annulus travel-time differences $[\delta t_{\rm{oi}}]$
from the linear simulation (blue with symbols) with the travel-time difference
computed from the ray theory with the same flow perturbations (red line).
The model used in the ray-theory computation is the convectively
stablized one used in the simulation.
The error bars are computed from the
scatter far from the feature locations.
No filtering has been done before the travel-time
measurements.  The average travel-time difference in the range
$\Delta=10$\,--\,$24^\circ$ has been scaled to match the observationally
determined mean
$5.1\secs$.  The same scaling factor is then used to scale the ray-theory
results.  
}
\label{F-sim_vert}
\end{figure}

In Article I, a convectively stabilized solar model \cite{Hanasoge06} 
was used with 
vertical-flow features with flow peaking at a depth of $z_0=-2.3\Mm$ 
with Gaussian 
depth profile with width $\sigma_z=0.82\Mm$ and horizontal Gaussian width
$\sigma_h=5.1\Mm$.  A global simulation of wave propagation is performed
with wave sources near the surface \cite{Hanasoge07}.  
Center to annulus travel-time
differences $[\delta t_{\rm{oi}}]$ were measured from the simulation results
as a function of the annulus radius $[\Delta]$.  To obtain travel times 
similar to the observed $5.1\secs$ required the peak amplitude of the
Gaussian flow to be $338\ms$.  Ray-theory calculations were made and we
found that the $\delta t_{\rm{oi}}$ for the ray theory were 24\,\% larger, 
suggesting some 
problem.  However, we had used the standard Model S \cite{Jcd96} 
to do the ray-theory
calculations where we should have been using the convectively stablilized
model, which it turns out makes a significant difference.  
A revised version of Figure 2 of Article I is shown in Figure~\ref{F-sim_vert}.
On the distance range $10$\,--\,$24 ^\circ$, the average measured travel-time 
difference $[\delta t_{\rm{oi}}]$ is now very close to the ray-theory
prediction.  This result shows that vertical flows like the ones suggested
in Article I are correctly modeled by the ray theory, as long as one is using
the correct background model.  It does not tell us, however, what might
happen to the corresponding horizontal flows and whether the separation of
horizontal and vertical flows by the $\delta t_{\rm{oi}}$ and
$\delta t_{\rm{qu}}$ is valid.

To test whether the horizontal- and vertical-flow components are separated
by the measurements of $\delta t_{\rm{oi}}$ and $\delta t_{\rm{qu}}$,
a simulation was done using a flow model identical to model {\sf g2} of 
Article I.
The solar model used is the convectively stabilized one described above
and a Cartesian simulation is done using the SPARC code 
(\opencite{Hanasoge06b},\citeyear{Hanasoge08};\opencite{Hanasoge06}).  500
features identical to the flow in model {\sf g2} are placed randomly in the
horizontal plane.  Center--annulus and quadrant travel-time differences
are measured as described above.  The horizontal spacing of the simulation
is $5.7\Mm$ with $512\times512$ pixels.  
The maximum depth is $104\Mm$.  The attempt was
to go as deeply as possible in order to be able to use large distances.
In order to be able to put the 500 features over the entire horizontal
field and still be able to obtain full annulus coverage for large separations, 
the horizontal periodicity of the simulation was used in the travel-time
computation.  

\begin{figure}
\centerline{\includegraphics[width=1.0\textwidth,clip=]{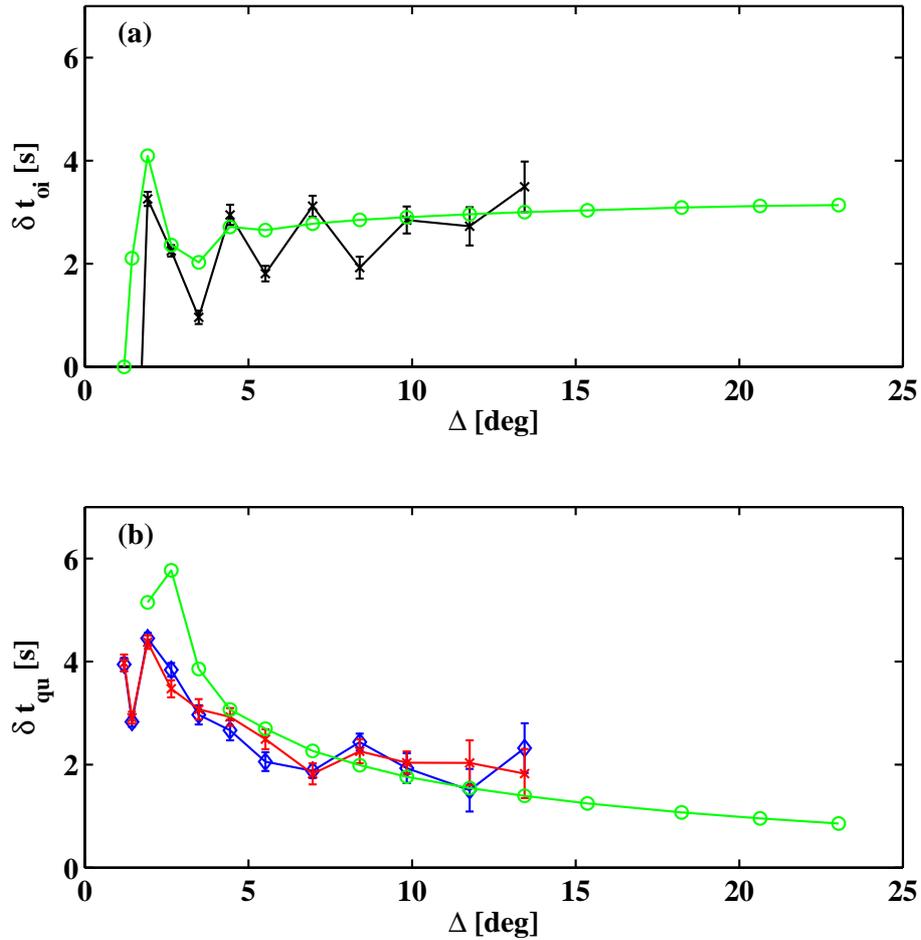}}
\caption{
Comparison of travel-time differences measured from a simulation with 
ray-theory computations.
(a) Center-annulus travel-time differences $[\delta t_{\rm{oi}}]$ from the
simulation (black with crosses, error bars, and connecting lines) and 
ray-theory computations of $\delta t_{\rm{oi}}$ (green circles). 
(b) Quadrant travel-time differences 
$\delta t_{\rm{we}}$ (blue diamonds) and 
$\delta t_{\rm{ns}}$ (red crosses) for the simulation and 
ray-theory computations for $\delta t_{\rm{we}}$ (green circles).
}
\label{F-sim_oiwe}
\end{figure}

The travel-time differences from the simulation and ray-theory computations
are shown in Figure~\ref{F-sim_oiwe}.  The
travel-time differences are only shown up to $\Delta=13^\circ$, as the
simulation was not deep enough to go to larger $\Delta$.  
The $\delta t_{\rm{oi}}$
measurements in Figure~\ref{F-sim_oiwe}a seem noisier than expected 
from the error bars.  There is
a rough agreement with the ray theory.  The asymptotic limit at 
$\Delta=24^\circ$ is only $3.1\secs$, as opposed to the expected $5.1\secs$.  
This is because in the convectively stabilized model the sound speed
is modified, which affects the ray-theory estimate of the travel-time
differences in the integral $\int{{\rm dr}(v/c^2)}$.
There is less general agreement of the quadrant travel-time differences
with the ray theory.  Especially for $\Delta<5^\circ$, the ray theory is
predicting too large a travel-time difference, while for $\Delta>10^\circ$
the opposite may be the case.

For both the spherical simulation (Figure~\ref{F-sim_vert}) and for the
Cartesian one (Figure~\ref{F-sim_oiwe}a), the ray theory has general 
agreement with the $\delta t_{\rm{oi}}$ measured from the simulations.
This suggests that the ray theory can adequately predict $\delta t_{\rm{oi}}$ 
for the range of $\Delta$ examined and for flows at this depth of
$z=-2.3\Mm$.  For the quadrant travel times, the situation is more 
problematic.  One question is whether a more shallow flow peaking at the
surface could somehow have its quadrant signal mimic the $\delta t_{\rm{oi}}$
signal of $5.1\secs$.  This will need to wait for future work.

\subsection{Travel Times Versus Temporal Frequency $[\nu]$}
\label{sec-nu}
\begin{figure}
\centerline{\includegraphics[width=1.0\textwidth,clip=]{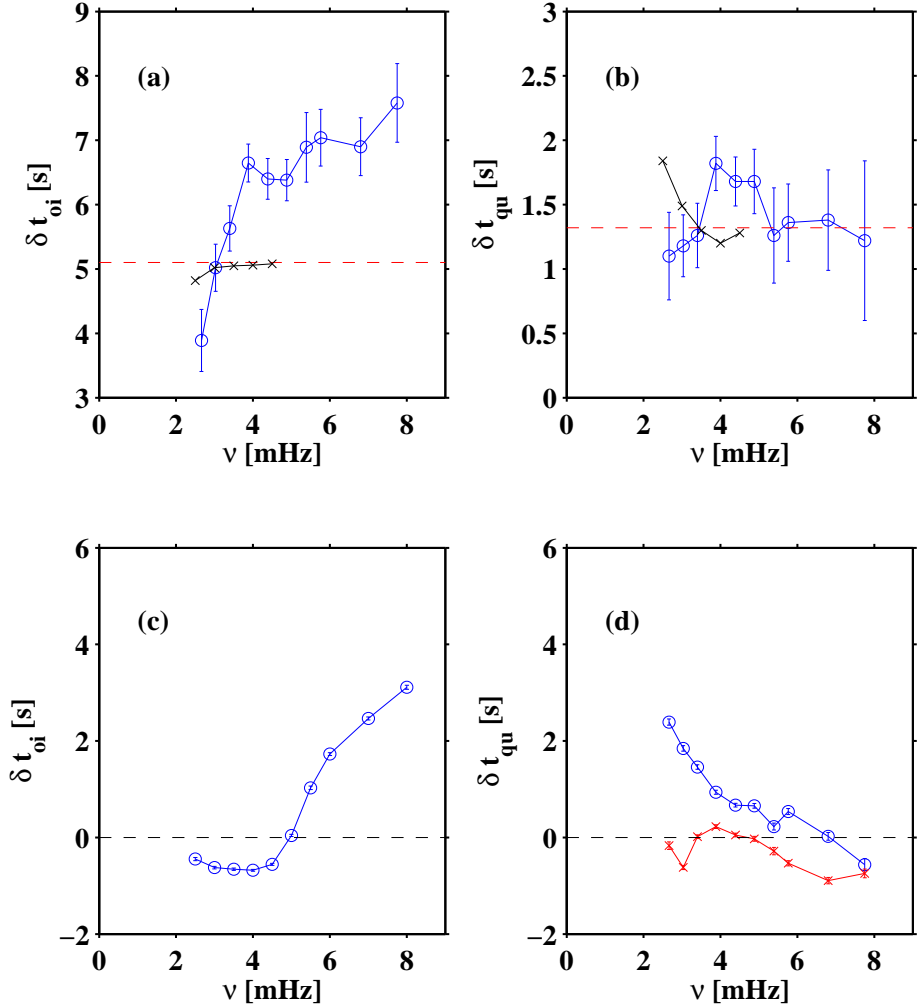}}
\caption{
Peak signals and backgrounds versus the temporal frequency $[\nu]$ for the
largest distance band $\Delta=22.1$\,--\,$24^\circ$. (a) The
center--annulus travel-time difference $[\delta t_{\rm{oi}}]$ versus the
temporal frequency $[\nu]$ (blue circles).  
The red horizontal dashed line is at $5.1\secs$, the result
for the $\nu$-averaged data.  
The black crosses and line are ray-path calculations based on model {\sf g2} 
from Article I.
(b) The peak quadrant travel-time difference
$[\delta t_{\rm{qu}}]$ (blue cirlces), where the
$\delta t_{\rm{we}}$ and $\delta t_{\rm{ns}}$ 
signals have been averaged to
reduce the error bar size.  The red horizontal dashed line is at the value for
the $\nu$-averaged result, $1.3\secs$.
The black crosses and curve are ray-path calculations based on model {\sf g2} 
from Article I.
(c) The center--annulus background signal far from
the supergranule centers versus $\nu$ (blue circles).  
This signal was subtracted from the
observed signal to yield the result in (a).  
The dashed black line indicates zero.
(d) The background signals
for $\delta t_{\rm{we}}$ (blue circles) and $\delta t_{\rm{ns}}$ (red crosses).
These were
subtracted from the observed signals and then the 
$\delta t_{\rm{we}}$ and $\delta t_{\rm{ns}}$ 
were averaged to yield
the result in (b).  
The dashed black line indicates zero.
}
\label{F-vs_freq}
\end{figure}

Another way to test the validity of the ray theory applied to flow measurements
is to measure the travel times versus the temporal frequency $[\nu]$.  To
first order, at the same distance $\Delta$, the travel-time differences due
to flows should be constant, according to the ray theory.  To our knowledge, 
such a test has not been carried out, although the frequency dependence 
of travel times has been measured extensively, e.g. \inlinecite{Dombroski13}.
For the ridge filtering used in that study, a $\nu$-dependence is expected,
however.
Such a test was conducted for the 64 12-hour intervals at the largest
distance range used, $\Delta=22.1$\,--\,$24^\circ$.
The travel-time  results versus $\nu$ are shown in Figure~\ref{F-vs_freq}.

To obtain these results, the datacubes were filtered as before with the 
phase-speed filter with constant degree width $\Gamma_\ell=400$, and in addition
a frequency filter.  Ten separate frequency filters were used.  The
first seven had central frequencies from $\nu=2.5\mHz$ to $\nu=5.5\mHz$ 
in steps of $0.5\mHz$ with power full width at half maximum of $0.5\mHz$.
The last three filters had central frequencies of $\nu=6,7,8\mHz$ with 
power full width at half maximum of $1.0\mHz$.
Center--annulus and quadrant cross-correlation maps were constructed for
each of the 64 12-hour intervals.  Average correlations were used to
measure guess times for the different $\nu$ intervals.  Gizon--Birch travel-time
maps were computed for the 64 intervals and the results were averaged
over $\Delta$ and the various differences computed.  Average maps about
the supergranulation centers were made.  Background signals were measured
far from the central feature for each $\nu$ for the different signals.
These are shown in Figure~\ref{F-vs_freq}c and \ref{F-vs_freq}d.  These 
are subtracted from the peak signals observed to yield the results
for $\delta t_{\rm{oi}}$ in Figure~\ref{F-vs_freq}a and for the two
quadrant signals subsequently averaged to yield Figure~\ref{F-vs_freq}b.

The center--annulus travel time differences $[\delta t_{\rm{oi}}]$ 
in Figure~\ref{F-vs_freq}a are clearly
not constant with frequency but increase by a factor of $\approx$two 
over the $\nu$ range.
For $\nu<\nu_{\rm ac}$, 
where $\nu_{\rm ac}\approx5\mHz$ is the peak acoustic-cutoff 
frequency of the atmosphere, the variation is approximately linear with a
zero intercept.  For $\nu>\nu_{\rm ac}$, the increase with frequency is 
somewhat smaller.  There have not been any predictions of flow travel times
for $\nu>\nu_{\rm ac}$, but it was feasible to measure them, and so it was done.
Ray-theory calculations of the travel-time differences are also shown in 
Figure~\ref{F-vs_freq}a and Figure~\ref{F-vs_freq}b.  
This result casts some doubt on the simple interpretation of the 
$\delta t_{\rm{oi}}$ measurements in terms of the ray theory. 
The acoustic-cutoff frequency used in the ray calculations is Lamb's
definition $\omega_c=c/2H_p$ \cite{Lamb09}, 
where $c$ is the sound speed and $H_p$ is the pressure scale height.  
It is interesting that the peak quadrant travel times $[\delta t_{\rm{qu}}]$
in Figure~\ref{F-vs_freq}b
are approximately constant over the $\nu$ range, although with a mean signal
of $1.3\secs$, the uncertainty in this conclusion is large.
The ray-theory estimates are consistent with the measurements for
the quadrant travel times.

The background signal for the $\delta t_{\rm{oi}}$ shown in 
Figure~\ref{F-vs_freq}c is very interesting.  
It is measured far from the central location, but is basically consistent
with taking the average signal over a large area.
In the trapped mode region
$[\nu<\nu_{\rm ac}]$, the signal is relatively small and negative, while for
$\nu>\nu_{\rm ac}$, it becomes positive and increases with $\nu$.
We have no particular hypothesis for the source of this signal, but this
additional information on the $\nu$ variation may be important for 
understanding it.  The background for the quadrant signals 
$[\delta t_{\rm{we}}$ and $\delta t_{\rm{ns}}]$ are shown in 
Figure~\ref{F-vs_freq}d.  
These signals are measured far from the central location in the North--South
direction for the $\delta t_{\rm{we}}$ and in the West--East direction for
the $\delta t_{\rm{ns}}$ signal.
The $\delta t_{\rm{we}}$ signal has a linear variation over the
frequency range.  This background signal is possibly due to solar rotation,
although it would imply that the solar rotation yields a $\nu$-dependent
$\delta t_{\rm{we}}$ while the supergranular signal does not.
Whatever causes background signals, 
they need to be removed from the supergranular signal.

\section{Conclusions}
\label{sec-dis}

The bulk of the evidence in the present article continues to support
a model of the average supergranulation cell as having an upflow 
with a velocity much larger than the surface upflow of $10\ms$, possibly
as large as $240\ms$ and a peak flow $2$\,--\,$3\Mm$ 
below the surface as seen in the best model {\sf g2}.  
In Article I, center--annulus travel time differences $[\delta t_{\rm{oi}}]$
were shown to agree well with model {\sf g2}, while in the present article,
the quadrant travel-time differences $[\delta t_{\rm{we}}$ and 
$\delta t_{\rm{ns}}]$ also agree mostly with this type of model.  However,
there is some disagreement that varies with $\Delta$ for the 
$\delta t_{\rm{qu}}$, suggesting that either the functional form of the
model needs to be adjusted or the ray kernels are incorrect, or both.

The apparent disagreement between the present work and the smaller flows
seen before has largely disappeared with the work of \inlinecite{Svanda12}.
That article 
does an analysis of average supergranules
similar to the averaging done in the present article.  He used $f$-modes and
small separation p-modes and finds flows that largely confirm the present
results.  There may be a factor of two difference between the two results,
which needs to be resolved.

The lack of a center-to-limb variation of the $\delta t_{\rm{oi}}$ signal
is useful for a general check of systematic errors.  
However, the $\nu$-dependence of the $\delta\tau_{\rm oi}$ signal 
as predicted by ray theory differs from observations, suggesting that
ray theory may be inaccurate in near-surface layers. The errors may 
be due to unmodeled finite-frequency effects or possibly differences 
in the acoustic-cutoff frequency between the Sun and Model S (used in
ray modelling here).  That the acoustic-cutoff frequency may differ
from that derived from Model S was shown by \inlinecite{Jefferies94}.
In a model with the reflection point that is a significant function of 
frequency,
the travel-time difference $[\delta t_{\rm{oi}}]$ could then also be 
a function of frequency.  
At a minimum, the variation in $\delta t_{\rm{oi}}$ by a factor of roughly
two over the $\nu$-range observed would seem to make the suggested flows
uncertain by a similar factor.

The simulation results (Section~\ref{sec-sim}) show that the type of
model considered (model {\sf g2}) does induce the kind of travel-time shifts
observed.  These are then seen by both the travel-time shifts measured
from the simulation and by ray theory calculated with the model used.
It is unfortunate that the modification to the solar model to stablize
it has such a large effect on the resulting $\delta t_{\rm{oi}}$.

Some of the earlier work finds the flow velocity peaking very near the surface
with a monotonic decrease with depth \cite{Birch06,Woodard07}.
These results would appear to be inconsistent with this article and 
Article I.  It was suggested in Article I that the perturbations to the
p-mode spectrum due to supergranulation are significant in $\ell$,
the spherical-harmonic degree.  The idea is that the supergranulation
pattern, with a spectrum peaking near $\ell=120$ would induce a
modulation of a $p$-mode ridge that would have a width in $\ell$ of
at least twice this amount.  To capture all of the supergranulation
signal requires a filter of at least a full-width-half-max of
$\Gamma_{\ell}=240$ and likely larger.  This justifies the value chosen for
the present work of $\Gamma_{\ell}=400$, which clearly captures all
of the $\delta t_{\rm{oi}}$ signal at large $\Delta$.  This conclusion
is supported by Figure 4 of Article I.  Of course, if the modeling is
correct, one can use any filter.  However, if much of the supergranulation
signal is not being captured, one becomes much more sensitive to the
modeling.  In \inlinecite{Woodard07}, the filters are 
narrower than used here, particularly at low frequencies.

Because the acoustic wavelength at the depth of the peak flow is larger
than the depth of the peak flow, 
ray theory may show inaccuracies. To 
improve the quality
of the flow model, it would therefore be useful to include finite-frequency 
effects \cite{Birch07}.  Further, the functional
dependence of travel times on the background-flow model may exceed
the linear limit if flow speeds are indeed on the order of $700\ms$.
Therefore a non-linear inversion for supergranular flow may be necessary to
explain the measured travel times \cite{Hanasoge11}.

%

%

%

%
%
\begin{acks}
The data used here are courtesy of NASA/SDO and the HMI Science Team.  We
thank the HMI team members for their hard work.  This work is supported by
NASA SDO. 
\end{acks}

%
%
%
%
%
%


\end{article} 
\end{document}